\documentclass[twocolumn,showpacs,preprintnumbers,amsmath,amssymb,prb,floatfix]{revtex4}

\usepackage{graphicx}
\usepackage{dcolumn}
\usepackage{bm}
\usepackage{epstopdf} 
 
\begin{document}
\bibliographystyle{prsty}    
 

\title{ On the Origin of the  2DEG Carrier Density at the LaAlO$_3$/SrTiO$_3$ 
Interface}
\author{Zoran S. Popovi\'c \cite{INN}, Sashi Satpathy\cite{MU}, and Richard M. Martin}
\affiliation{Department of Physics, University of Illinois at Urbana-Champaign, 1110 W. Green St., Urbana, IL 61801, USA
}
\date{\today}

\begin{abstract}

Transport measurements of the two-dimensional electron gas (2DEG) at the LaAlO$_3$/SrTiO$_3$
interface 
have found a density of carriers much lower than expected from the ``polar catastrophe" arguments.
From a detail density-functional study, we suggest how this discrepancy may be reconciled.  We find that 
electrons occupy multiple subbands at the interface leading to
a rich array of transport properties.  Some electrons are confined to a single interfacial layer
and susceptible to localization, while others with small masses and extended over several layers are expected to contribute to transport.

 \end{abstract}

\maketitle

\begin{figure}
\centering
\includegraphics[width=7.0cm, angle=0]{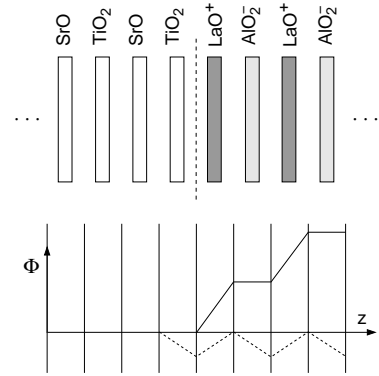}  
\caption{Nominal layer charges at the n-type interface with LaO/TiO$_2$ termination. Bottom figure shows 
the diverging Coulomb potential (``polar catastrophe", solid line) and an example of how it may be avoided by adding half an electron to the single TiO$_2$ layer at the interface (dashed line). 
}
\label{structure}
\end{figure}

In a seminal paper,  Ohtomo and Hwang\cite{ohtomo} reported the observation of a conducting layer of electrons  at the interface between two insulators LaAlO$_3$ (LAO) and SrTiO$_3$ (STO). The carrier density was quite high, of the order of $ 10^{17}$ cm$^{-2}$,
as also the mobility, leading to the prospect for high-performance oxide electronics. 
Consequently, considerable attention has been focused on the understanding 
of the origin of these charge carriers.\cite{eckstein} 
Systematic experiments on oxygen annealed samples as well as on samples grown under high oxygen pressure conditions have tuned down the carrier density to about $ 1-2  \times 10^{13}$ cm$^{-2}$, which is now the common carrier density observed by 
several groups.\cite{brinkman,kalabukov,beasley,fert,thiel}  
Thus the emerging consensus is that the higher carrier density $ \sim 10^{17}$ cm$^{-2}$, originally observed, is  due to the electrons released by the oxygen vacancies and they have a 3D character, while the lower carrier density 
$ \sim 10^{13}$ cm$^{-2}$ may be an intrinsic feature of the interface. 
However, this intrinsic carrier density
is much smaller than the 0.5 electrons per unit cell ($ 3.5 \times 10^{14}$ cm$^{-2}$) needed to suppress the 
``polar catastrophe,"\cite{harrison-cat} 
which arises because of the diverging
Coulomb field due to the polarity of the individual layers
(Fig. \ref{structure}).
Thus the origin of the carrier density for the intrinsic interface is poorly understood and remains a key puzzle for these interfaces. 

In this paper, we suggest
a solution for this puzzle based on the results of our detail density-functional study of the electronic structure. Although there also exists a second type of interface, viz., the p-type interface with SrO/AlO$_2$ termination, we focus on the n-type interface in this paper which is illustrated in Fig. 1. Our calculations reveal the presence of different types of 
bands at the interface. We argue that some of these electrons may become localized at the interface due to disorder or electron-phonon coupling and may not participate in the conduction, while the others form delocalized states contributing to transport.


Density functional calculations were performed using the
generalized gradient approximation (GGA) for the exchange and correlation.\cite{GGA}
 We have not included the Coulomb corrections within the LDA+U functional, as it may not describe the correlation effects properly for low carrier density situation, away from half filling. 
The Kohn-Sham equations were solved 
by either the linear augmented plane waves (LAPW)\cite{wien2k} or the linear muffin-tin orbitals (LMTO) methods\cite{andersen}.
Earlier density-functional studies have been reported in
the literature\cite{pentcheva,freeman,albina,hellberg}; 
however, since the supercells used in these studies were quite small, the nature of the
interface states could not be determined, which is the primary focus of our paper to develop our arguments.

\begin{figure}
\centering
\includegraphics[width=6cm]{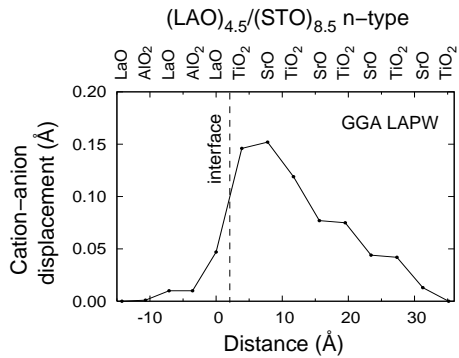} 
\caption{Computed relaxed atomic positions, consistent with the spread of the interfacial electrons into many TiO$_2$ layers. 
}
\label{relaxed-structure}
\end{figure}

We used the supercell (STO)$_{7.5}$/(LAO)$_{7.5}$ for studying the interface electron states, where the half layers in the formula signify an extra layer of TiO$_2$ or LaO, so that we have two identical n-type interfaces
 (TiO$_2$/LaO) present in the
structure. The atomic relaxation was performed using the LAPW method for a slightly different supercell,
viz., for (STO)$_{8.5}$/(LAO)$_{4.5}$ to keep the calculations manageable. This is reasonable since results presented in Fig. \ref{relaxed-structure} show  very little relaxation in the LAO part beyond the second layer, which in turn is consistent with the
small charge leakage to the LAO side. 
 
The GGA optimized bulk lattice 
constants are 3.94 \AA\ and 3.79 \AA\, for the STO and the LAO bulk, respectively,
which are quite close to their experimental values (3.905 and 3.789 \AA). For the optimization, we fixed the lattice constant for the Sr part to be 3.91 \AA, both
in the directions parallel and normal to the interface
as appropriate for samples grown on the SrTiO$_3$ substrate,
while the lattice constant for LAO normal to the interface was chosen
to preserve the volume of the bulk unit cell. This defined the supercell 
which was held fixed, while the internal positions of the atoms were allowed to
change in a series of self-consistent calculations until the forces became smaller
than about 6 mRy/ au. The main effect of the relaxation,
shown in Fig. \ref{relaxed-structure}, is to polarize the cation and the anion planes near the interface similar to the relaxation for the
well-studied LaTiO$_3$/SrTiO$_3$ interface.\cite{popovic,okamoto1,hamann,popovic-relaxed}
For the present interface, the atomic polarization occurs mostly in the STO part, with the oxygen atoms moving 
towards the interface, while the cations move away from it, so that there is a buckling in the planes.
The largest cation-anion polarization of about 0.15 \AA\ occurs for the interfacial TiO$_2$ and the SrO layers
as seen from Fig. \ref{relaxed-structure}.  

\begin{figure}
\centering
\includegraphics[width=6.0cm]{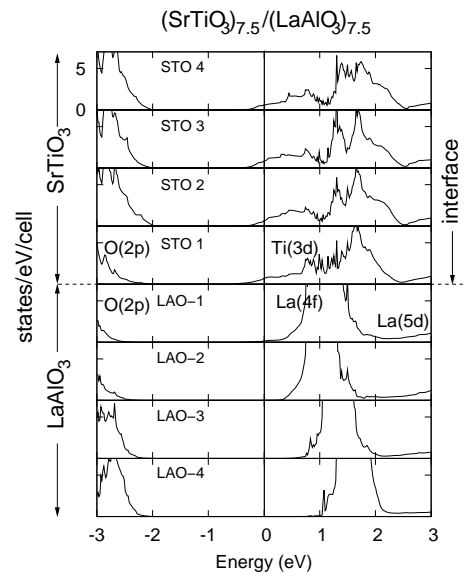} 
\caption{Layer-projected densities-of-states indicating the spread of the 
interface electrons (0.5 e$^-$/cell occupying Ti(d) bands) into several layers near the interface on the STO side. 
The Fermi energy is taken as zero.
}
\label{ldos}
\end{figure}

\begin{figure}
\centering
\includegraphics[width=3.2cm]{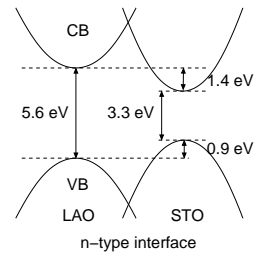}
\caption{Computed  band offset  at the interface. Valence band positions were computed by using the O-1s core state as the reference level. The conduction band edges were inferred by using the experimental band gaps.\cite{sto-lao-bandgap}  
}
\label{offset}
\end{figure}

\begin{figure*}
\centering
\includegraphics[height=4.0cm, angle=0]{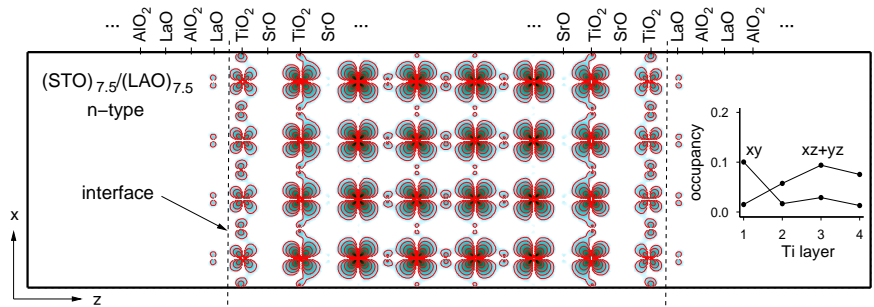}
\caption{(Color online) Charge contours of the occupied states in the conduction bands (Fig.\ref{bands}) plotted on the xz-plane (normal to the interface). Electrons penetrate very little to the LAO side.  Note that only the xz orbitals, but {\it not} the xy or yz orbitals, show up on such a plane. Nevertheless, the contours faithfully reproduce the spread of some of the electrons (see text) deeper into the STO part. The starting density for the 
contours is $3.2 \times 10^{-5}$e/\AA$^3$ and successive values increase by the factor of 3. Inset shows the Ti (d) occupancy as a function of the distance of the Ti layers from the interface.} 
\label{contour}
\end{figure*}

The layer-projected densities-of-states (Fig. \ref{ldos}) calculated with the LMTO method shows that the half electron per unit cell
needed to overcome the polar catastrophe resides on the STO side
and spreads out to several Ti layers near the interface. 
The shift of the band edges from layer to layer indicates an electric field discontinuity  consistent with the accumulation of a positive monopole charge at the interface. Electrons attracted to this monopole  reside on the STO side, as expected from  the computed band line-up  at the interface (Fig. \ref{offset}) and confirmed from the
charge density contour plot (Fig. \ref{contour}).

 
\begin{figure}
\centering
\includegraphics[width=6.5cm]{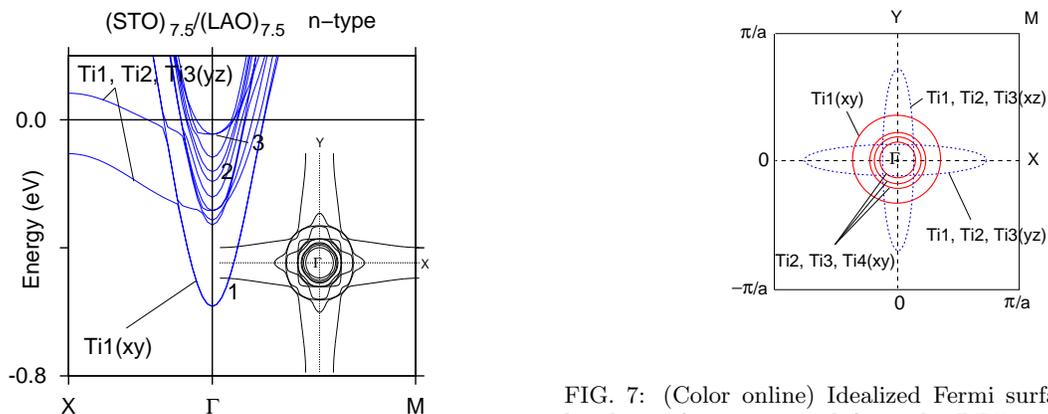}
\caption{Conduction bands near   the Fermi energy
and their predominant orbital characters.  Band dispersions are shown along the interface (xy plane)
with $X = \pi/a (1,0,0)$ and $M = \pi/a (1,1,0)$, where `a' is the in-plane lattice constant. 
}
\label{bands}
\end{figure}

An important result of this work is the existence of
two types of carriers at the interface with possibly different transport
properties, which is inferred by examining the  
conduction band structure
presented in Fig. \ref{bands}.  
The occupied conduction bands are made out of Ti(d) t$_{2g}$ states. 
As indicated from the wave function characters of Table I,
the lowest band is well localized on the first Ti layer because of a much lower interface potential there
(Recall that the polar catastrophe argument
would place as many as 0.5 electrons per cell at the interface).
The xy bands are lower than the yz or zx bands because of the larger hopping integrals
along both directions in the xy plane, parallel to the interface.
Above these bands at the $\Gamma$ point are several other bands which occur
close together in energy. These are xy-type bands located on the second or the third Ti layer and the yz or xz-type bands, the latter being made out of a linear combination of states on several Ti layers and have small dispersions along one of the two planar directions. Apart from the lowest 
band, states in the remaining bands spread several layers into the STO part and because of the finite size supercell used in our calculations, the bands belonging to the two interfaces mix,
producing a splitting of the energies, visible, e.g., near the X point in the figure. This
particular split becomes amplified in the Fermi surface, producing the 
two outermost wings both along X and Y directions (inset of Fig. \ref{bands}). 
The two outermost wings of the Fermi surface would not 
split without this interaction. 


\begin{table}[b]
\centering
\hsize=8cm
\caption{Square of the wave function characters for typical conduction states at the $\Gamma$ point,
labelled in Fig. \ref{bands}. The TiO$_2$ planes have been numbered consecutively starting from the interface.}
\vspace{14pt}
 \begin{tabular}{cccccc}
 \hline
\multicolumn{1}{c}{State}& (TiO$_2)_1$& (TiO$_2)_2$ & (TiO$_2)_3$& (TiO$_2)_4$& rest\\ 
\hline
\multicolumn{1}{c}{$\psi_1$}&0.82 & 0.02 & 0 & 0 & 0.16\\
\multicolumn{1}{c}{$\psi_2$}&0.02 & 0.17 & 0.44 & 0.26 & 0.11\\
\multicolumn{1}{c}{$\psi_3$}&0.01 & 0.35 & 0.18 & 0.37 & 0.09\\
\hline
\end{tabular}
\end{table}

\begin{figure} 
\centering
\includegraphics[width=4.5cm]{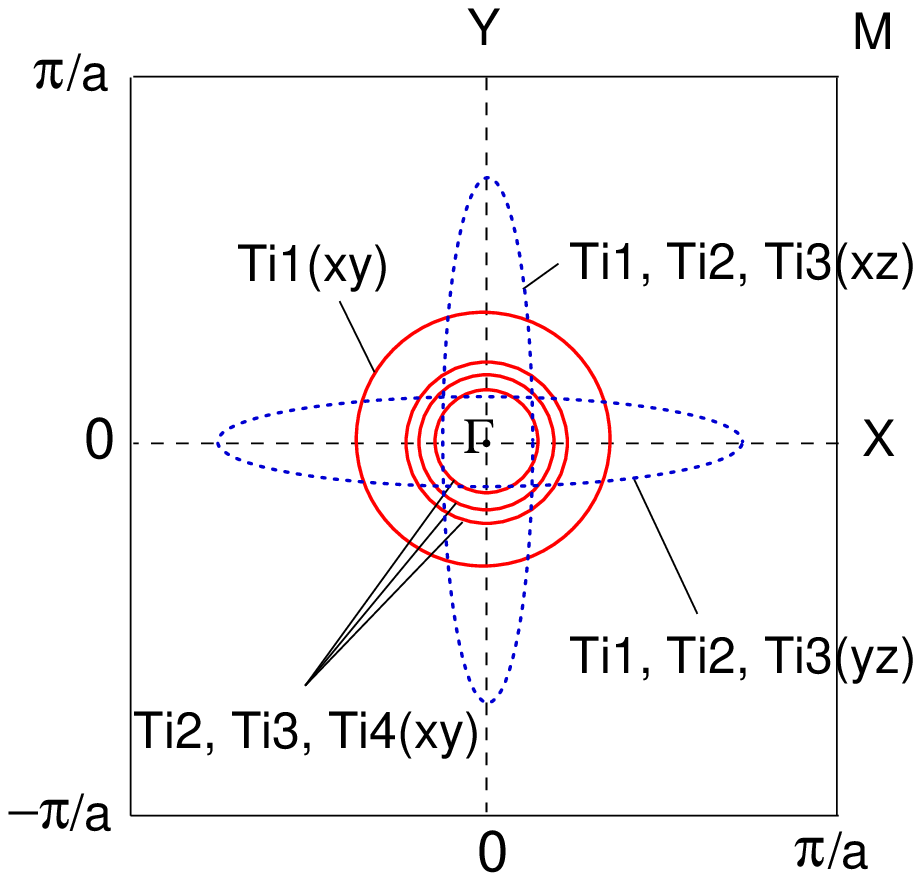}
\caption{(Color online) Idealized Fermi surface for the isolated interface, extracted 
from the DFT results of Fig. \ref{bands}. 
Hybridization between bands rounds out the 
sharp crossing points in the Fermi surface, 
leading to more complex orbits as seen in the inset of
Fig. \ref{bands}. 
 }
\label{fermiideal}
\end{figure}

The basic features of the Fermi surface for the isolated interface is illustrated in Fig. \ref{fermiideal}, which is extracted from the calculated Fermi surface of Fig. \ref{bands} and where, 
for pedagogical reasons, we have omitted the hybridization between the bands. The circles are derived from the xy orbitals, which disperse with the tight-binding matrix element of $V_{dd\pi}$ along
both $\hat{x}$ and $\hat{y}$, resulting in the tight-binding band structure: 
$\epsilon^{(xy)}(k)=2 V_{dd\pi} (\cos k_x a + \cos k_y a)$, which is isotropic for small Bloch momenta,
while
 the xz and the yz orbitals disperse with the tight-binding matrix element $V_{dd\pi}$ along
one direction but with a much smaller\cite{harrison}  $V_{dd\delta}$ along the other, so that we have for the xz bands:
$\epsilon^{(xz)}(k)=2 V_{dd\pi} \cos k_x a + 2 V_{dd\delta} \cos k_y a$, and similarly for $\epsilon^{(yz)}(k)$, which produce the elliptic Fermi surfaces.


Our results suggest a possible explanation for the
observed low carrier density for the samples grown under oxygen rich conditions. This growth condition is thought to produce an intrinsic interface, where
the observed carrier density is  $ \sim 1-2 \times 10^{13}$ cm$^{-2}$, corresponding to only about 0.015-0.03 electrons per interface unit cell, much smaller than the half electrons per cell needed to suppress the polar catastrophe. These electrons  need not reside on just one interfacial layer, rather, they can be spread over several layers and the polar catastrophe is avoided so long as the total electron count is right. Not all of these electrons might contribute to transport.

From Fig. \ref{fermiideal}
we infer the presence of two types of electrons at the interface.  The first 
originates from the lowest conduction subband, which has a strong 2D character consisting mostly of Ti(d) states on the first
layer
and forming the largest circle
in the Femi surface plot.
It is well-known that all states in 2D are Anderson localized by 
disorder and these states may therefore become localized.
In addition, these electrons may form self-trapped polarons due to the Jahn-Teller coupling of the Ti(d$^1$) center. The unpaired electron could then behave as a Kondo center. 
The  xz/yz states which form the elliptic Fermi circles in Fig. \ref{fermiideal}  
 are also prone to localization on account of their heavy masses along a planar direction.
All these  electrons may therefore not participate in transport. 

The second type of electrons are from the Ti (xy) bands above the lowest band which spread over several Ti layers into the bulk and show up as the  three innermost circles in the Fermi surface plot. 
These electrons have small masses along the plane, 
	they spread several Ti layers into the bulk and are as a result difficult to localize
	by disorder or electron-phonon coupling
	and we argue  that they are likely the mobile electrons that would
	contribute to transport  in the intrinsic samples. 
Since the  electron density in the Ti(d) bands is low, far from half filling, electron correlation effects are not expected to alter the metallic behavior apart from a mass renormalization in a Fermi liquid picture. 
	
	The number of these mobile electrons, estimated from the area of the three innermost circles in Fig. \ref{fermiideal} is about  $8 \times10^{13}$  cm$^{-2}$, which is 
	 five times smaller than the electrons needed to satisfy the polar catastrophe. It is however still several times higher than  the measured carrier density  in the intrinsic samples. 
	 	This discrepancy could be due to several reasons. 
		Density-functional calculations tend to yield too high an energy for 
localized states, which would mean that the density in the delocalized bands is overestimated.
Another possibility is that the  mobile carrier density may be reduced due to additional localization effects
not considered here explicitly. So, according to this scenario, the mobile electrons arise from the inner pockets in the Fermi surface, while the remaining electrons are localized.

	There are  indeed signatures of these two types of carriers in the experiments. For instance, 
	Brinkman et al.\cite{brinkman} were able to explain their transport data in terms of two types of
	carriers.  These experiments showed a Kondo resistance minimum as a function of temperature,
	which they interpreted in terms of itinerant electrons scattering off of localized Ti (3d) moments.
	Within our scenario,
	the formation of the local moments is entirely likely due to the localization of some of the electrons.
	The delocalized electrons lead to the possibility of superconductivity.\cite{reyren}
	
	It would be valuable to measure the Hall effect on the intrinsic samples, where 
	several authors have been unsuccessful in observing the Shubnikov-de Haas (SdH) 
	 oscillations. It is possible that the oscillations become
	 suppressed due to the intermixing of the localized and the itinerant parts of the 
	 Fermi surface, leading to strong scattering between them before the completion of full magnetic orbits.
	 In that case, one might have to go to higher magnetic fields, where magnetic tunneling between the bands might lead to complete orbits.

We thank Jim Eckstein and Michael Ma for insightful  discussions and B. R. K. Nanda for help with the computations. ZP and SS would like to acknowledge the hospitality of the Materials Computation Center at the University of Illinois. This work was supported by the U. S. Department of Energy through Grant No.
DE-FG02-00ER45818.



\begin{thebibliography}{00}


\bibitem[*]{INN}
Permanent address: INN-``Vin\v ca'',
PO Box: 522, 11001 Belgrade, Serbia 

\bibitem[$\dagger$]{MU}
Permanent address: Department of Physics, University of Missouri, Columbia, MO 65211, USA



\bibitem{ohtomo} A. Ohtomo and H. Y. Hwang, Nature (London) {\bf 427}, 423 (2004)

\bibitem{eckstein} J. N. Eckstein, Nature Materials {\bf 6}, 473 (2007)

\bibitem{brinkman} A. Brinkman {\it et al.}, 
Nature Materials {\bf 6}, 493 (2007)

\bibitem{beasley} W. Siemons
{\it et al.} Phys. Rev. Lett. {\bf 98}, 196802 (2007)

\bibitem{kalabukov} A. Kalabukov {\it et al.} Phys. Rev. B {\bf 75}, 121404 (2007)

\bibitem{fert} G. Herranz et al., Phys. Rev. Lett. {\bf  98}, 216 803 (2007)


\bibitem{thiel}  S. Thiel {\it et al.}, 
Science {\bf 313}, 1942 (2006)

\bibitem{harrison-cat} W. A. Harrison {\it et al.}, 
Phys. Rev. B {\bf 18}, 4402 (1978); E. A. Kraut, 
Phys. Rev. B {\bf 31}, 6875 (1985)

\bibitem{GGA} J. P. Perdew and Y. Wang,
Phys. Rev. B {\bf 45}, 13244 (1992); 
J. P. Perdew {\it et al.},
Phys. Rev. Lett. {\bf 77}, 3865 (1996)

\bibitem{wien2k} P. Blaha {\it et al.}, 
WIEN2k, An Augmented Plane Wave + Local Orbitals Program 
for Calculating Crystal Properties (Karlheinz Schwarz, Techn. 
Universit\"{a}t Wien, Austria), 2001. ISBN 3-9501031-1-2 

\bibitem{andersen} O. K. Andersen and O. Jepsen, Phys. Rev. Lett 
{\bf 53}, 2571 (1984)


\bibitem{pentcheva} R. Pentcheva and W. E. Pickett, Phys. Rev. B {\bf 74}, 035112 (2006)

\bibitem{freeman} M. S. Park, S. H. Rhim, and A. J. Freeman, Phys. Rev. B {\bf 74},
205416 (2006)

\bibitem{albina} J.-M. Albina {\it et al.}, 
Phys. Rev. B {\bf 76}, 165103 (2007) 

\bibitem{hellberg} C. Cen {\it et al.}, Nat. Mater. {\bf 7}, 298 (2008)

\bibitem{popovic} Z. S. Popovic and S. Satpathy, Phys. Rev. Lett. 
{\bf 94}, 176805 (2005)

\bibitem{okamoto1} S. Okamoto, A. J. Millis, and N.A. Spaldin, Phys. Rev. 
Lett. {\bf 97}, 056802 (2006)  

\bibitem{hamann} D. R. Hamann, D. A. Muller, and H. Y. Hwang, Phys. Rev. B 
{\bf 73}, 195403 (2006)


\bibitem{popovic-relaxed} P. Larson, Z. S. Popovic and S. Satpathy, Phys. Rev. B
{\bf 77}, 245122 (2008)

\bibitem{sto-lao-bandgap} J. A. Moland, Phys. Rev. {\bf 94}, 724 (1954);
M. Capizzi and M. Frova, Phys. Rev. Lett. {\bf 25}, 1298 (1970);
S. G. Lim {\it et al.}, J. Appl. Phys. {\bf 91}, 4500 (2002);
K. van Benthem and C. Els\" asser, J. Appl. Phys. {\bf 90}, 6156 (2001)

\bibitem{harrison}W. A. Harrison, \textit{Electronic Structure and the Properties of
Solids} (Freeman, San Francisco, 1980)



\bibitem{reyren}   N. Reyren, {\it et al.}, 
Science {\bf 317}, 1196 (2007)



\end{thebibliography}
\end{document}